\documentclass[aps,prb,showpacs,12pt]{revtex4-1}
\usepackage{graphicx}
\newcommand{\hide}[1]{}

\begin{document}
\title{Ferroelectricity in strained Ca$_{0.5}$Sr$_{0.5}$TiO$_3$ from first principles}
\author{Christopher R. Ashman}
\address{1602 W. 42$^{nd}$ St., Richmond, Virginia 23225  U.S.A.}
\author{C. Stephen Hellberg}
\address{Center for Computational Materials Science Code 6393, Naval Research Laboratory, Washington D.C., 20375 U.S.A.}
\author{Samed Halilov}
\address{Department of Materials Science and Engineering, MIT, 77 Mass. Ave., Cambridge, Ma. 02139  U.S.A.}

\begin{abstract}
We present a density functional theory investigation of 
strained
Ca$_{0.5}$Sr$_{0.5}$TiO$_3$.  We have
determined the structure and polarization for a number of arrangements of Ca and Sr in a 2$\times$2$\times$2
supercell.  The a and b lattice vectors are strained to match the lattice constants of the 
rotated Si(001) face.  To set the context for the CSTO study, we also include simulations of the 
Si(001) constrained structures for CaTiO$_3$ and SrTiO$_3$.  Our primary findings are that all 
Ca$_{0.5}$Sr$_{0.5}$TiO$_3$ structures examined except one are ferroelectric, exhibiting polarizations 
ranging from 0.08 C/m$^2$ for the lowest energy configuration to about 0.26 C/m$^2$ for the higher energy 
configurations.  We find that the configurations with larger polarizations have lower c/a ratios.  
The net polarization of the cell is the result of Ti-O ferroelectric displacements regulated by  A-site
cations.  
\end{abstract}
\maketitle
%
%======================================================================
%\chapter*{Fe$_3$S$_4$ data }
%\section*{09-06-06}
%======================================================================
%
There is great interest in combining ferroelectrics with semiconductors.
Potential devices include nonvolatile memory, reprogrammable logic,
and even quantum computation\cite{lev01}.
Ideally, ferroelectrics would be grown on the technologically dominant Si(001) surface,
and there has been great effort at growing SrTiO$_3$ on 
Si(001)\cite{mck98,woi06,woi07,fit08,lev09}.
The in-plane lattice parameters of SrTiO$_3$ are 
1.7\% larger than those of the (110) and ({\-1}10) directions of the Si(001) surface,
and
thin films of compressively strained STO have been shown to exhibit the desired ferroelectric behavior\cite{woi07,lev09}.
However,
as the thickness of the film increases, the STO relaxes to the unstrained state and the ferroelectric behavior is 
no longer observed\cite{woi06,lev09}.  One possibility to induce ferroelectricity into SrTiO$_3$ films is to introduce 
dopants such as either Ba or Ca\cite{kle00,bed84,gen08}.  For example Ca$_{(1-x)}$Sr$_x$TiO$_3$ offers the 
possibility of invoking ferroelectricity in SrTiO$_3$ while at the same time maintaining a closer
lattice 
match to the Si(001) surface.  Although there have been a number of experimental studies aimed at identifying 
the crystal structure and lattice properties of unstrained Ca$_{(1-x)}$Sr$_x$TiO$_3$ \cite{car06,woo06,hui07}, 
to date the authors are not aware of any theoretical studies of the strained case.

In this work we present a theoretical study of ferroelectrically induced polarization
in bulk, strained Ca$_{0.5}$Sr$_{0.5}$TiO$_3$.  Structural optimizations were done using the VASP
code\cite{vasp1,vasp2} with the Projector Augmented Wave method to treat the electronic structure
problem\cite{paw}.  The gradient corrected XC-functionals were treated within the Perdew, Becke 
and Ernzerhof scheme \cite{pbe96}.  In all cases we used a supercell consisting of $2\times2\times2$ 
primitive cells or two layers of four primitive cells each.  This requires eight alkaline-earth 
atoms in the unit cell while allowing for the inclusion of rotation and tilting effects in the oxygen 
octahedra.  Periodic boundary conditions were employed to simulate a spatially extended material.  We 
used a $4\times4\times4$ k-point mesh and a planewave cutoff energy of 350 eV.  The unit cell was  
strained to match the lattice constants of the 45$^{\circ}$ rotated Si(001) face.  For our unit cell 
the {\bf a} and {\bf b} lattice vectors are 7.728 \AA .   The {\bf c} lattice vector is obtained by 
calculating the total energy for a range of values separated by 0.01 \AA\ approximately centered on the 
cubic value until a minimum energy configuration was bracketed.  For the total energy calculations the 
lattice parameters were frozen but the atomic positions were free to relax until the forces were smaller 
than 0.005 eV/\AA\ along any cartesian direction for any atom.  

For the calculation of polarization we used the Abinit code\cite{abinit} and FHI98 pseudopotentials of 
the Trouiller-Martins type\cite{fhi99}.  The structure was imported directly from the VASP code structural 
optimization step and no relaxation was performed for the lattice or ions.  

\begin{figure}[t!]
\includegraphics[angle=0,width=16cm,draft=false]{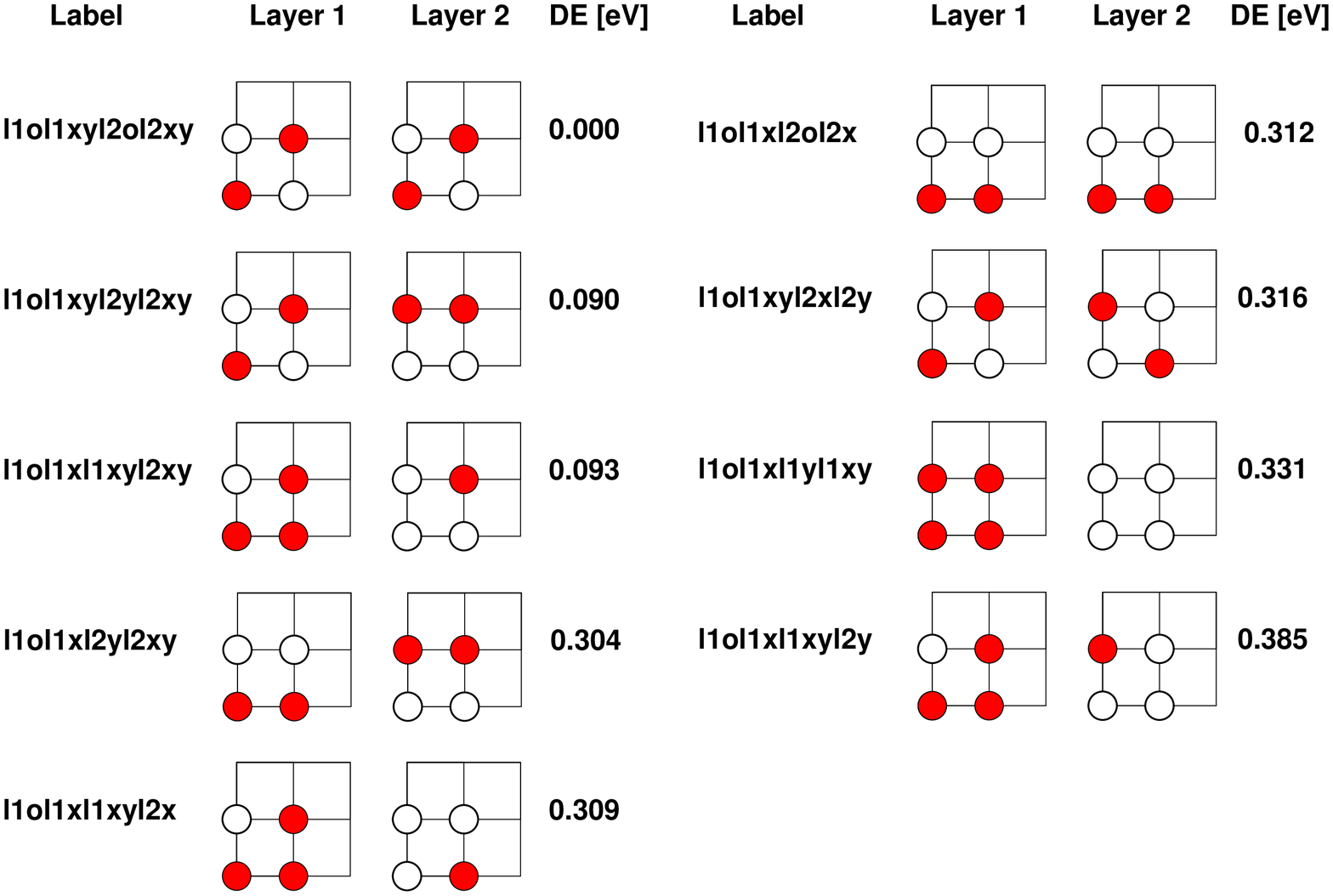}
 \caption{ (Color online) Top view schematic of the configurations considered in this work.  The left 
           column is an ID for each configuration, the second column is a schematic of 
           the first layer, the third column is a schematic of the second layer.  The
           final column gives the relative energies. The white/red circles represent 
           the positions of the Ca/Sr atoms on the xy-plane lattice face.  
 }
 \label{fig1}
 \end{figure}

In the chosen unit cell there are 8 A-sites (alkaline-earth) of which four are occupied by Ca and four by Sr.  
Most of the 70 possible arrangements of the cations are related by rotational and mirror symmetries.
We examined the 9 configurations shown in Fig.\ \ref{fig1}.

Our primary result is that all the configurations we considered except one exhibit a net ferroelectric 
dipole moment.  The lowest-energy configuration consists of alternating stripes of Ca and Sr running 
diagonally across the strained face of the perovskite.  This configuration possesses a net polarization 
of 0.077 C/m$^2$.  We find that there is a trend for the lowest energy configurations to have the smallest 
polarizations and the largest c/a ratio's.  We summarize this information in table \ref{tab3}.

\begin{table}[h!]
\begin{center}
\begin{tabular}{|cccc|}
\hline
\multicolumn{4}{|c|}{Ca$_{\frac{1}{2}}$Sr$_{\frac{1}{2}}$TiO$_3$} \\
\hline
Configuration \hspace{.5cm} &  $\Delta$E[eV]\hspace{.5cm} &  $\frac{c}{a}$ ratio \hspace{.5cm}& Polarization [C/m$^2$] \\
\hline
l1ol1xyl2ol2xy &   0.000   & 1.043       & 0.077 \\
l1ol1xyl2yl2xy &   0.090   & 1.048       & 0.035 \\
l1ol1xl1xyl2xy &   0.094   & 1.048       & 0.037 \\
l1ol1xl2yl2xy  &   0.304   & 1.020       & 0.000 \\
l1ol1xl1xyl2x  &   0.309   & 1.030       & 0.250 \\
l1ol1xl2ol2x   &   0.312   & 1.030       & 0.245 \\
l1ol1xyl2xl2y  &   0.316   & 1.033       & 0.253 \\
l1ol1xl1yl1xy  &   0.331   & 1.030       & 0.268 \\
l1ol1xl1xyl2y  &   0.385   & 1.027       & 0.031 \\
\hline
\end{tabular}
\caption{Tabulated data for the configurations at 50\% Ca, 50\% Sr pictured in fig.
         \ref{fig1}.  Included are the configuration identification, the total energy, 
         relative energy and polarization for the lowest energy of each configuration.
}
\label{tab3}
\end{center}
\end{table}
Several studies on SrTiO$_3$ have shown that quantum fluctuations may compete with ferroelectric distortions, leaving the
perovskite in an unpolarized state referred to as an incipient ferroelectric\cite{mul79,zho96}.  To address this possibility in
the doped perovskite, we have calculated the energy difference between the polarized and unpolarized cells for the two cases 
corresponding to the lowest energy configuration (l1ol1xyl2ol2xy) and the alternating stacked sequence of Ca and Sr (l1ol1xl1yl1xy).  
We find the energy differences to be 0.07 eV and 0.03 eV per formula unit (5 atom cell) for the lowest energy configuration and the
stacked configuration respectively.  If we use formula (1) from reference\cite{zho96} to estimate the contribution of
quantum fluctuations we come up with about 0.003 eV/oxygen or about .01 eV/formula unit.   Thus it is likely that 
the polarization in these cells persists.
 
\begin{figure}[t!]
\includegraphics[angle=-90,width=12cm,draft=false]{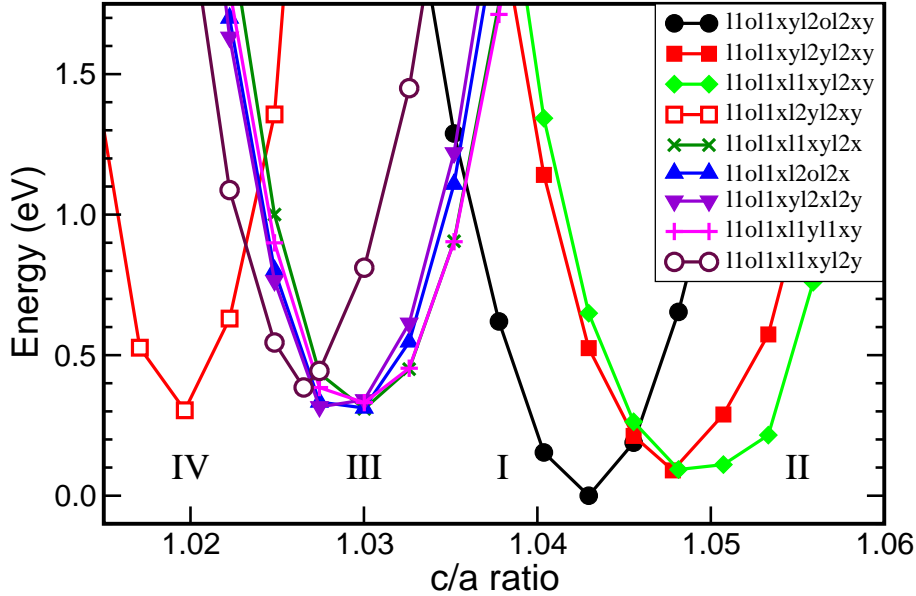}
 \caption{ (Color online) Plot of the relative energy as a function of c/a ratio for the configurations considered
           in this work.  The curves were shifted so that the globally lowest energy configuration 
           aligned with 0.00 eV.  The labels I, II, III, and IV correspond to arrangements of 
           a$^+$a$^+$c$^-$, a$^+$a$^+$c$^0$, a$^-$b$^+$c$^-$, and a$^-$b$^+$c$^+$ tilting 
           and rotation of the oxygen octahedra. 
 }
 \label{fig2}
 \end{figure}

\begin{figure}[t!]
\includegraphics[angle=0,width=10cm,draft=false]{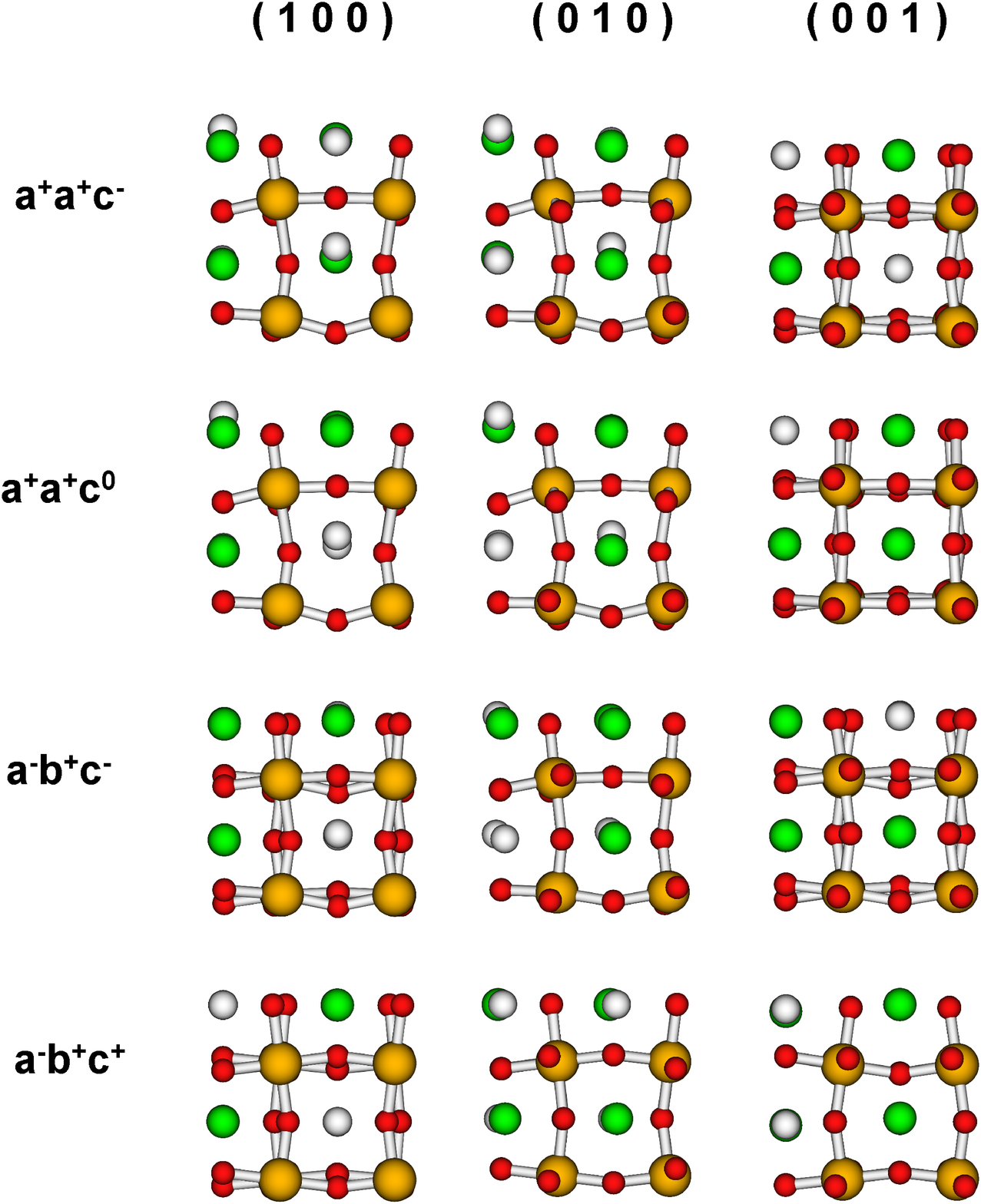}
 \caption{ Ball and Stick models of the lowest energy configuration corresponding to each combination
           of AFD rotations and tilting of the oxygen octahedra exhibited by configurations investigated
           in this work.   We present views down the (100), (010), and (001) planes respectively.  
           The models correspond to the configurations lying at the bottoms of the curves labeled 
           in figure \ref{fig1}.  In order from smallest to largest balls, the O are red, Ca white
           Sr green, and Ti yellow.
 }
 \label{fig3}
 \end{figure}

We find that polarization tends to increase with decreasing c/a ratio. There is an exception 
to this trend for the configuration labeled l1ol1xl2yl2xy which has no net polarization but has 
the smallest c/a ratio of those considered.  We have plotted the relative energies of the 
configurations as a function of c/a ratio in figure \ref{fig2}.  In this figure we label four 
families of curves which are differentiated based upon the AFD rotations and/or tiltings.  The 
four distinct combinations of rotation and tilting are presented in
figure \ref{fig3} which consist of projections down the (100), (010), and (001) faces of the
l1ol1xyl2ol2xy, l1ol1xyl2yl2xy, l1ol1xl1xyl2x, and l1ol1xl2yl2xy configurations respectively.  
For simplicity of expressing the rotations and tilting of the oxygen octahedra we employ the 
Glazer notation \cite{gla72}.  This notation was developed for ABX$_3$ perovskites and assumes rigid 
rotations of the oxygen octahedral cages whereas we are considering the case of AA$'$TiO$_3$ and allow 
distortions of the oxygen cages.  

Analysis of the displacement patterns shows that the tetragonal strain is driven by the AFD octahedral
rotations hybridized with the ferroelectric displacements (FE) rather than by the latter alone.  Earlier 
theoretical work has shown that the introduction of dopants with smaller atomic radii into the A-sites is
able to drive the tetragonal strain \cite{hal02}.  Thus the distribution of Ca-cations determines the rotation
of the oxygen cages versus the distortion along the tetragonal (c) axis.  On the other hand, the polarization 
is clearly associated with the FE displacements.  We find that FE 
displacements of the Ti atoms relative to the oxygen cages do not vary widely despite the range of polarizations 
for the different configurations.  For example, the Ti atoms undergo an average FE displacement 
from their centrosymmetric positions in the oxygen octahedra of 0.213 \AA\ for the l1ol1xyl2ol2xy configuration.  This configuration
has a polarization of 0.077 C/m$^2$, whereas the l1ol1xl1yl1xy configuration Ti atoms have an average FE 
displacement of 0.134 \AA\ while exhibiting the one of the largest net polarizations of 0.268 C/m$^2$.  Instead, we find 
that displacements of the A-site cations plays a significant role in determining the overall polarization of the cell.  
In the l1ol1xyl2ol2xy configuration, the A-site Ca cations have very large displacements along the c-axis.
The displacement of these cations is such that the Ca-O dipoles are anti-aligned with the Ti-associated polarization 
of the cell.  If the Ca ions are shifted to be coplanar with the oxygen atoms in the same atomic layer thus removing 
their contribution to the polarization in the cell, the polarization increases from 0.077 C/m$^2$ to 0.186 C/m$^2$.  
If all A-site cations are shifted to their nonpolar locations, the polarization increases further to 0.222 C/m$^2$.  
In their optimal locations, the Ca-ions have large displacements of up to 0.589-0.650 \AA\ in the three lowest energy configurations.
Other configurations have Ca-ion displacements of 0.000 to 0.243 \AA.  The resulting picture is such that the 
configurations with the lowest energy, groups I and II in Fig.~\ref{fig3}, have the highest tetragonal strain 
and lowest polarization as featured by the relatively strong a$^+$a$^+$-type octahedral rotations.  These rotations
will be referred to as tiltings hereafter in the text. 

We consider the effect of the position within the cell of the atomic arrangements on the stability and polarization 
by examining the three cases of the lowest energy l1ol1xyl2ol2xy configuration, the l1ol1xyl2xl2y 
configuration which may be thought of as a rock-salt type structure, and finally the l1ol1xl1yl1xy configuration which is 
constructed of alternating planes of Ca and Sr ions in the (001) plane.  Note that for the l1ol1xyl2ol2xy and l1ol1xyl2xl2y 
configurations, the sequence of the A and A$'$ cations (A = Ca, A$'$ = Sr) along the ({\bf a}, {\bf b}) lattice directions is A-A$'$- 
and so on, while that of the l1ol1xl1yl1xy is either A-A- or A$'$-A$'$-.  Along the {\bf c} lattice direction, the l1ol1xyl2ol2xy 
give either a stacking of A-A- or A$'$-A$'$- whereas the other two give A-A$'$- and so on.  Considering these cations as hard spheres, 
the packing of A-A$'$- is more compact than that of A$'$-A$'$- in the ideal tetragonal structure, since the Ca ion is smaller than the 
Sr ion.  We have confirmed that the l1ol1xyl2ol2xy and l1ol1xyl2xl2y undergo smaller compressive strains of (0.5\%,0.5\%),
and (0.9\%, 0.7\%) compared to (1.2\%, 0.7\%) for the l1ol1xl1yl1xy configuration along the {\bf a} and {\bf b} lattice 
directions.  This results in reduced polar AFD rotations.  For the l1ol1xyl2ol2xy configuration, the average octahedral rotation about
the tetragonal axis is 5.7$^\circ$, while the l1ol1xl1yl1xy configuration has an average octahedral rotation of about 7.7$^\circ$.
In the {\bf c} direction, both the l1ol1xyl2xl2y and l1ol1xl1yl1xy are more compact  than the l1ol1xyl2ol2xy which restricts 
the range of displacements that the Ca ions are permitted.  This in turn reduces their ability to respond to the internal 
electric field established by FE distortions.  

  To assist in understanding how doping affects the stability
  and polarization of the perovskite we have performed a series of
  calculations in which we include only the effects of FE displacements,
  FE displacements coupled with AFD rotations, and FE displacements coupled
  with octahedral tilting about one axis.  The results are presented in
  table \ref{tab4}.  We find that freezing out the tilting and allowing only 
  AFD rotations of the octahedra coincides with an increase in the cell polarization 
  relative to the case where only FE displacements are allowed, contrary to what 
  is found for STO\cite{sai00}. This behavior is the result of large $\sim$0.5 
  \AA\ displacements of the Ca atoms out of the (001) plane of the oxygens in 
  the case where the AFD distortions are suppressed.  The Ca atoms are displaced 
  such that the Ca associated electric dipole opposes the contribution from the 
  Ti associated polarization.  This is facilitated by a large c/a ratio for the 
  structure.  When AFD rotations are included the strain on the octahedral cages 
  is reduced and the tetragonal distortion is smaller.  In this case the Ca atom 
  displacements are reduced to about 0.1 \AA\  due to the steric hindrance of the 
  oxygen atoms.  Thus their associated electric dipole is of a smaller magnitude.  

  The inclusion of tilting without AFD rotations also acts to reduce the
  cell polarization.   Due to the compressive strain imposed in the plane,
  the oxygen octahedra are compressed and there is a
  large tetragonal distortion out of plane which allows for displacement
  of the Ca atoms.  The distance between the Ca atoms and the nearest 
  oxygen atoms in the Ti-O plane above them (relative to oxygen octahedra that
  are displaced downwards due to polarization effects) is about 2.43 \AA\
  as compared to the sum of the ionic radii of Ca and O which is about
  2.38-2.42 \AA\ assuming Shannon ionic radii \cite{sha76}.  Thus they
  are bounded in their displacement primarily by the size of the O and
  Ca ionic shells.  For comparison, the sum of the ionic radii of Sr and
  O is 2.56-2.60 \AA .  Generally speaking, the A-site cations exhibit
  distances to nearest neighbor oxygens from adjacent Ti-O planes that
  correspond to within 0.05 \AA\ of the sum of their ionic radii.

\begin{figure}[t!]
\includegraphics[angle=0,width=10cm,draft=false]{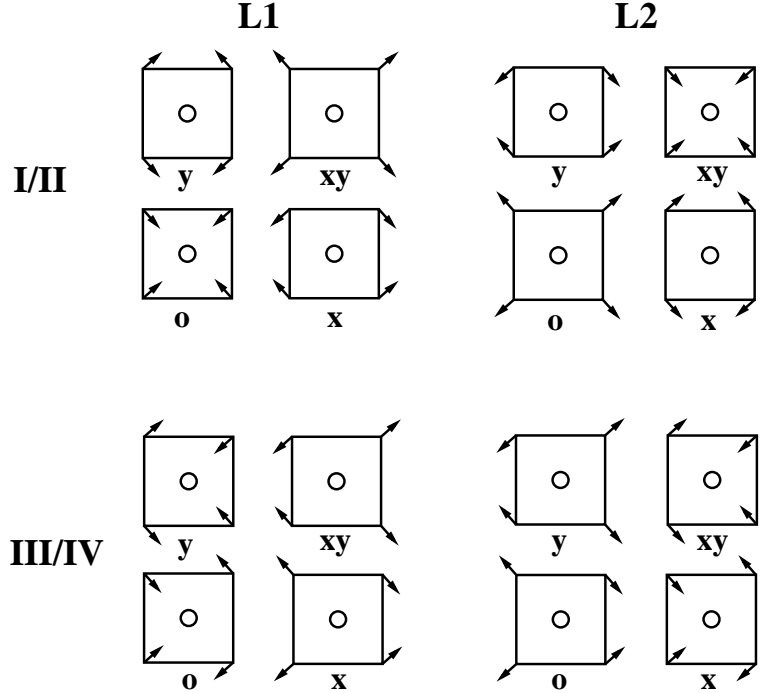}
 \caption{ 
  Schematic diagrams of the (001)-plane projected tilting of the oxygens
  surrounding the A-site cations.  The top set of eight panels show the 
  direction of oxygen displacements corresponding to curves I and II of
  figure \ref{fig2}.  The bottom set correspond to those of curves III
  and IV.  In each case, the left 4 panels are for layer 1 (L1), the 
  right set are for layer 2 (L2).  Each schematic is labeled with an o,
  x, y, or xy to indicate its proximate placement in the unit cell.
  Cells adjacent to each other share oxygen atoms and direction vectors
  indicate that. 
 }
 \label{fig4}
 \end{figure}

  Based on these observations, the placement of the Ca and Sr atoms in the
  lattice can be explained.  Fig. \ref{fig4} consists of two sets of schematics
  indicating only the directions of the displacements of the A-site planar oxygen
  atoms relative to the A-site cations due to tilting of the octahedral cages.
  Note that the four sets of curves in fig. \ref{fig2} reduce to the two sets 
  of oxygen displacement schematics (including rotation about the c-axis gives 
  the complete 4 sets of curves).  A study of the figure indicates that once 
  the displacement of the oxygens about one cation have been established, the 
  displacements of all oxygens in the unit cell is specified, assuming rigid 
  oxygen octahedra.  

  In layer 1 (L1) of the fig. \ref{fig4} top set of schematics, the schematic
  labeled "o" indicates that all of the nearest neighbor oxygens that are
  co-planar with the A-site cation are tilted towards the cation.  This set
  of tiltings corresponds to the large out-of-plane displacements that have been
  observed for the Ca atoms in the simulations.  The lowest energy configurations
  contain this tilting arrangement.  This set of tiltings, coupled with 
  the large A-site cation out-of-plane displacements, is associated with the 
  lowest energy configurations.  In all 9 configurations studied here, no such 
  set of tiltings is associated with the Sr atoms due to their large ionic radii.
  The tilting drives the A-site cations out of the plane.  In order for the Sr
  atoms to displace, the tetragonal distortion must increase with a corresponding
  increase in strain energy.  On the other hand the Ca atoms
  are able to displace with no additional contribution to the
  tetragonal strain energy due to their smaller atomic radii.  To quantify this,
  note that, in the l1ol1xyl2ol2xy configuration, the O-O nearest neighbor
  separation in A-site plane is 3.29 \AA .  The in-plane distance from one of
  the oxygens to the center of the square face where a non-displaced A-site
  cation would reside is 2.33 \AA .  In order to accommodate a Ca ion, given
  the ionic radii of the Ca and O, the Ca ion would need to be displaced by
  about 0.59 \AA\ which is the same as the observed Ca displacement.  In order
  to accommodate an Sr cation, the tetragonal distortion would need to be 
  increased by 0.53 \AA\ per layer.

  Given the importance of this octahedral tilting arrangement we are able to
  explain why the Ca atoms align along (110) planes as opposed
  to a rock-salt type configuration.  Observation of the schematics I/II
  in fig. \ref{fig4} indicates that there can be only one such tilting
  configuration per layer.  These configurations must be staggered diagonally
  in order not to impose large distortions on the oxygen octahedra.  Such
  an arrangement excludes the rock salt configuration.  It also excludes
  alternating planes composed of Ca and Sr cations.  

\begin{table}[h!]
\begin{center}
\begin{tabular}{|cccccccccccc|}
\hline
               &\multicolumn{2}{ c }{Tetragonal Unpoled} &\multicolumn{3}{ c }{Tetragonal Poled} & \multicolumn{3}{ c }{AFD Only} & \multicolumn{3}{ c|}{Tilting Only} \\
Config. &  $\delta$E[eV] &  $\frac{c}{a}$ &  $\delta$E[eV] &  $\frac{c}{a}$ & $\mu$ [C/m$^2$] &  $\delta$E[eV] &  $\frac{c}{a}$ & $\mu$ [C/m$^2$] &  $\delta$E[eV] &  $\frac{c}{a}$ & $\mu$ [C/m$^2$] \\
\hline
A  & 1.438  & 1.025  & 0.968 & 1.067 & 0.090 & 0.639 & 1.042 & 0.250  & 0.756  & 1.053  & 0.020  \\
B  & 1.442  & 1.025  & 1.034 & 1.067 & 0.087 & 0.646 & 1.030 & 0.164  & 0.749  & 1.056  & 0.030  \\
C  & 1.352  & 1.025  & 1.058 & 1.061 & 0.035 & 0.647 & 1.030 & 0.162  & 0.633  & 1.052  & 0.070  \\ 
\hline
\end{tabular}
\caption{Tabulated data for the A=l1ol1xyl2ol2xy, B=l1ol1xyl2xl2y and C=l1ol1xl1yl1xy configurations at 50\% Ca, 50\% Sr pictured in fig.
         \ref{fig1}.  The first two columns are the difference in energy between the given configuration and the l1ol1xyl2ol2xy ground state
         (relative energy) and the c/a ratio for the configurations with no rotations 
         or tilting and no FE displacements of the atoms in the unit cell.  Each of the subsequent sets of three columns
         give the relative energy, c/a ratio and cell polarization with only FE displacements, AFD rotations 
         and tilting about the a-axis of the cell.
}
\label{tab4}
\end{center}
\end{table}

In table \ref{tab1} we compare the lattice parameters and polarizations for the Si(001) lattice matched CaTiO$_3$ and SrTiO$_3$ with the
lowest energy configuration of Ca$_{0.5}$Sr$_{0.5}$TiO$_3$.  The c/a ratios sandwich the Ca$_{0.5}$Sr$_{0.5}$TiO$_3$ 
value of 1.043 with 1.002 and 1.053 respectively and with respect to the 40 atom cell.  We find that the mixed configuration has the lowest 
net polarization. Both Ca- and SrTiO$_3$ are considered incipient ferroelectrics in their unstrained states but have been found to be polar in
their strained configurations \cite{ekl09,hae04}.  The oxygen cage rotations may be classified as a$^+$b$^-$c$^-$ and a$^0$a$^0$c$^-$ for the CaTiO$_3$ and 
SrTiO$_3$ respectively.  In these pure ABO$_3$ perovskites the A-site cations have smaller FE displacements than in the mixed case. 
  
\begin{table}[t!]
\begin{center}
\begin{tabular}{|lcccc|}
%\hline
%\multicolumn{5}{|c|}{CaTiO$_3$ and SrTiO$_3$ lattice vectors and polarization} \\
\hline
\multicolumn{1}{|c}{Configuration} \hspace{.5cm} &  {\bf a} (\AA ) \hspace{.5cm} &  {\bf b} (\AA ) \hspace{.5cm} & {\bf c} (\AA ) \hspace{.5cm}& Polarization [C/m$^2$] \\
\hline
Ca$_{0.5}$Sr$_{0.5}$TiO$_3$      & 7.728  \hspace{.5cm} & 7.728  \hspace{.5cm} & 8.06             & 0.077 \\
CaTiO$_3$                        & 7.728  \hspace{.5cm} & 7.728  \hspace{.5cm} & 7.740            & 0.146 \\
SrTiO$_3$                        & 7.728  \hspace{.5cm} & 7.728  \hspace{.5cm} & 8.136            & 0.304 \\
\hline
\end{tabular}
\caption{
Lattice vectors and polarization for Si lattice matched Ca$_{0.5}$Sr$_{0.5}$TiO$_3$ (lowest energy configuration), CaTiO$_3$ and SrTiO$_3$ 
as determined in this work.
}
\label{tab1}
\end{center}
\end{table}
 
In summary we have identified a ground state configuration for the Si(001) lattice matched Ca$_{0.5}$Sr$_{0.5}$TiO$_3$.  
This consists of alternating planes of Ca and Sr running along the (110) direction relative to the cubic ABO$_3$ 
perovskite phase.  We find this structure to be ferroelectric with a net polarization of 0.077 C/m$^2$ and a reduced 
surface strain compared to STO.  We have identified eight higher energy configurations which, with one exception, 
exhibit net ferroelectric polarizations.  The A-site cations counter the dominant ferroelectric contributions of 
the Ti atoms.  This work shows that the small ionic radius of the Ca-ions combined with octahedral rotations allows 
them to displace in such a way as to minimize the net cell polarization.  The placement of the A-site cations 
determines whether AFD rotations or tilting effects are dominant.  The former suppresses octahedral distortions 
and subsequent displacement of the A-site cations.  When tilting is dominant, the tetragonal distortion is enhanced 
and the cell polarization decreases.   

The present study does not include surface and interface effects which very likely play an important role in both 
the overall structure and polarization of thin film structures.  However, it identifies the key interactions which govern the
stability and polarization of AA$'$TiO$_3$ perovskites which should prove valuable in designing such materials. 

Computations were performed at the AFRL DoD Major Shared Resource Center.
%

%====================================================================
\bibliography{./sto}

%merlin.mbs 2010-03-15 4.21a (PWD, AO, DPC)
\begin{thebibliography}{26}%
\makeatletter
\providecommand \@ifxundefined [1]{%
 \@ifx{#1\undefined}
}%
\providecommand \@ifnum [1]{%
 \ifnum #1\expandafter \@firstoftwo
 \else \expandafter \@secondoftwo
 \fi
}%
\providecommand \@ifx [1]{%
 \ifx #1\expandafter \@firstoftwo
 \else \expandafter \@secondoftwo
 \fi
}%
\providecommand \natexlab [1]{#1}%
\providecommand \enquote  [1]{``#1''}%
\providecommand \bibnamefont  [1]{#1}%
\providecommand \bibfnamefont [1]{#1}%
\providecommand \citenamefont [1]{#1}%
\providecommand \href@noop [0]{\@secondoftwo}%
\providecommand \href [0]{\begingroup \@sanitize@url \@href}%
\providecommand \@href[1]{\@@startlink{#1}\@@href}%
\providecommand \@@href[1]{\endgroup#1\@@endlink}%
\providecommand \@sanitize@url [0]{\catcode `\\12\catcode `\$12\catcode
  `\&12\catcode `\#12\catcode `\^12\catcode `\_12\catcode `\%12\relax}%
\providecommand \@@startlink[1]{}%
\providecommand \@@endlink[0]{}%
\providecommand \url  [0]{\begingroup\@sanitize@url \@url }%
\providecommand \@url [1]{\endgroup\@href {#1}{\urlprefix }}%
\providecommand \urlprefix  [0]{URL }%
\providecommand \Eprint [0]{\href }%
\@ifxundefined \urlstyle {%
  \providecommand \doi  [0]{\begingroup \@sanitize@url \@doi}%
  \providecommand \@doi [1]{\endgroup \@@startlink {\doibase
  #1}doi:\discretionary {}{}{}#1\@@endlink }%
}{%
  \providecommand \doi  [0]{doi:\discretionary{}{}{}\begingroup
  \urlstyle{rm}\Url }%
}%
\providecommand \doibase [0]{http://dx.doi.org/}%
\providecommand \Doi [0]{\begingroup \@sanitize@url \@Doi }%
\providecommand \@Doi  [1]{\endgroup\@@startlink{\doibase#1}\@@Doi}%
\providecommand \@@Doi [1]{#1\@@endlink}%
\providecommand \selectlanguage [0]{\@gobble}%
\providecommand \bibinfo  [0]{\@secondoftwo}%
\providecommand \bibfield  [0]{\@secondoftwo}%
\providecommand \translation [1]{[#1]}%
\providecommand \BibitemOpen [0]{}%
\providecommand \bibitemStop [0]{}%
\providecommand \bibitemNoStop [0]{.\EOS\space}%
\providecommand \EOS [0]{\spacefactor3000\relax}%
\providecommand \BibitemShut  [1]{\csname bibitem#1\endcsname}%
%</preamble>
\bibitem [{\citenamefont {Levy}(2002)}]{lev01}%
  \BibitemOpen
  \bibfield  {author} {\bibinfo {author} {\bibfnamefont {J.}~\bibnamefont
  {Levy}},\ }\href@noop {} {\bibfield  {journal} {\bibinfo  {journal} {Phys.
  Stat. Sol. B},\ }\textbf {\bibinfo {volume} {233}},\ \bibinfo {pages} {467}
  (\bibinfo {year} {2002})}\BibitemShut {NoStop}%
\bibitem [{\citenamefont {McKee}\ \emph {et~al.}(1998)\citenamefont {McKee},
  \citenamefont {Walker},\ and\ \citenamefont {Chisholm}}]{mck98}%
  \BibitemOpen
  \bibfield  {author} {\bibinfo {author} {\bibfnamefont {R.~A.}\ \bibnamefont
  {McKee}}, \bibinfo {author} {\bibfnamefont {F.~J.}\ \bibnamefont {Walker}}, \
  and\ \bibinfo {author} {\bibfnamefont {M.~F.}\ \bibnamefont {Chisholm}},\
  }\href@noop {} {\bibfield  {journal} {\bibinfo  {journal} {Phys. Rev.
  Lett.},\ }\textbf {\bibinfo {volume} {81}},\ \bibinfo {pages} {3014}
  (\bibinfo {year} {1998})}\BibitemShut {NoStop}%
\bibitem [{\citenamefont {Woicik}\ \emph {et~al.}(2006)\citenamefont {Woicik},
  \citenamefont {Li.}, \citenamefont {Zschack}, \citenamefont {Karapetrova},
  \citenamefont {Ryan}, \citenamefont {Ashman},\ and\ \citenamefont
  {Hellberg}}]{woi06}%
  \BibitemOpen
  \bibfield  {author} {\bibinfo {author} {\bibfnamefont {J.~C.}\ \bibnamefont
  {Woicik}}, \bibinfo {author} {\bibfnamefont {H.}~\bibnamefont {Li.}},
  \bibinfo {author} {\bibfnamefont {P.}~\bibnamefont {Zschack}}, \bibinfo
  {author} {\bibfnamefont {E.}~\bibnamefont {Karapetrova}}, \bibinfo {author}
  {\bibfnamefont {P.}~\bibnamefont {Ryan}}, \bibinfo {author} {\bibfnamefont
  {C.~R.}\ \bibnamefont {Ashman}}, \ and\ \bibinfo {author} {\bibfnamefont
  {C.~S.}\ \bibnamefont {Hellberg}},\ }\href@noop {} {\bibfield  {journal}
  {\bibinfo  {journal} {Phys. Rev. B},\ }\textbf {\bibinfo {volume} {73}},\
  \bibinfo {pages} {024112} (\bibinfo {year} {2006})}\BibitemShut {NoStop}%
\bibitem [{\citenamefont {Woicik}\ \emph {et~al.}(2007)\citenamefont {Woicik},
  \citenamefont {Shirley}, \citenamefont {Hellberg}, \citenamefont {Anderson},
  \citenamefont {Sambasivan}, \citenamefont {Fischer}, \citenamefont {Chapman},
  \citenamefont {Stern}, \citenamefont {Ryan}, \citenamefont {Ederer},\ and\
  \citenamefont {Li}}]{woi07}%
  \BibitemOpen
  \bibfield  {author} {\bibinfo {author} {\bibfnamefont {J.~C.}\ \bibnamefont
  {Woicik}}, \bibinfo {author} {\bibfnamefont {E.~L.}\ \bibnamefont {Shirley}},
  \bibinfo {author} {\bibfnamefont {C.~S.}\ \bibnamefont {Hellberg}}, \bibinfo
  {author} {\bibfnamefont {K.~E.}\ \bibnamefont {Anderson}}, \bibinfo {author}
  {\bibfnamefont {S.}~\bibnamefont {Sambasivan}}, \bibinfo {author}
  {\bibfnamefont {D.~A.}\ \bibnamefont {Fischer}}, \bibinfo {author}
  {\bibfnamefont {B.~D.}\ \bibnamefont {Chapman}}, \bibinfo {author}
  {\bibfnamefont {E.~A.}\ \bibnamefont {Stern}}, \bibinfo {author}
  {\bibfnamefont {P.}~\bibnamefont {Ryan}}, \bibinfo {author} {\bibfnamefont
  {D.~L.}\ \bibnamefont {Ederer}}, \ and\ \bibinfo {author} {\bibfnamefont
  {H.}~\bibnamefont {Li}},\ }\Doi {10.1103/PhysRevB.75.140103} {\bibfield
  {journal} {\bibinfo  {journal} {Phys. Rev. B},\ }\textbf {\bibinfo {volume}
  {75}},\ \bibinfo {eid} {140103} (\bibinfo {year} {2007})}\BibitemShut
  {NoStop}%
\bibitem [{\citenamefont {Kourkoutis}\ \emph {et~al.}(2008)\citenamefont
  {Kourkoutis}, \citenamefont {Hellberg}, \citenamefont {Vaithyanathan},
  \citenamefont {Li}, \citenamefont {Parker}, \citenamefont {Andersen},
  \citenamefont {Schlom},\ and\ \citenamefont {Muller}}]{fit08}%
  \BibitemOpen
  \bibfield  {author} {\bibinfo {author} {\bibfnamefont {L.~F.}\ \bibnamefont
  {Kourkoutis}}, \bibinfo {author} {\bibfnamefont {C.~S.}\ \bibnamefont
  {Hellberg}}, \bibinfo {author} {\bibfnamefont {V.}~\bibnamefont
  {Vaithyanathan}}, \bibinfo {author} {\bibfnamefont {H.}~\bibnamefont {Li}},
  \bibinfo {author} {\bibfnamefont {M.~K.}\ \bibnamefont {Parker}}, \bibinfo
  {author} {\bibfnamefont {K.~E.}\ \bibnamefont {Andersen}}, \bibinfo {author}
  {\bibfnamefont {D.~G.}\ \bibnamefont {Schlom}}, \ and\ \bibinfo {author}
  {\bibfnamefont {D.~A.}\ \bibnamefont {Muller}},\ }\Doi
  {10.1103/PhysRevLett.100.036101} {\bibfield  {journal} {\bibinfo  {journal}
  {Phys. Rev. Lett.},\ }\textbf {\bibinfo {volume} {100}},\ \bibinfo {eid}
  {036101} (\bibinfo {year} {2008})}\BibitemShut {NoStop}%
\bibitem [{\citenamefont {Warusawithana}\ \emph {et~al.}(2009)\citenamefont
  {Warusawithana}, \citenamefont {Cen}, \citenamefont {Sleasman}, \citenamefont
  {Woicik}, \citenamefont {Li}, \citenamefont {Kourkoutis}, \citenamefont
  {Klug}, \citenamefont {Li}, \citenamefont {Ryan}, \citenamefont {Wang},
  \citenamefont {Bedzyk}, \citenamefont {Muller}, \citenamefont {Chen},
  \citenamefont {Levy},\ and\ \citenamefont {Schlom}}]{lev09}%
  \BibitemOpen
  \bibfield  {author} {\bibinfo {author} {\bibfnamefont {M.~P.}\ \bibnamefont
  {Warusawithana}}, \bibinfo {author} {\bibfnamefont {C.}~\bibnamefont {Cen}},
  \bibinfo {author} {\bibfnamefont {C.~R.}\ \bibnamefont {Sleasman}}, \bibinfo
  {author} {\bibfnamefont {J.~C.}\ \bibnamefont {Woicik}}, \bibinfo {author}
  {\bibfnamefont {Y.}~\bibnamefont {Li}}, \bibinfo {author} {\bibfnamefont
  {L.~F.}\ \bibnamefont {Kourkoutis}}, \bibinfo {author} {\bibfnamefont
  {J.~A.}\ \bibnamefont {Klug}}, \bibinfo {author} {\bibfnamefont
  {H.}~\bibnamefont {Li}}, \bibinfo {author} {\bibfnamefont {P.}~\bibnamefont
  {Ryan}}, \bibinfo {author} {\bibfnamefont {L.-P.}\ \bibnamefont {Wang}},
  \bibinfo {author} {\bibfnamefont {M.}~\bibnamefont {Bedzyk}}, \bibinfo
  {author} {\bibfnamefont {D.~A.}\ \bibnamefont {Muller}}, \bibinfo {author}
  {\bibfnamefont {L.-Q.}\ \bibnamefont {Chen}}, \bibinfo {author}
  {\bibfnamefont {J.}~\bibnamefont {Levy}}, \ and\ \bibinfo {author}
  {\bibfnamefont {D.}~\bibnamefont {Schlom}},\ }\href@noop {} {\bibfield
  {journal} {\bibinfo  {journal} {Science},\ }\textbf {\bibinfo {volume}
  {324}},\ \bibinfo {pages} {367} (\bibinfo {year} {2009})}\BibitemShut
  {NoStop}%
\bibitem [{\citenamefont {Kleemann}\ \emph {et~al.}(2000)\citenamefont
  {Kleemann}, \citenamefont {Dec}, \citenamefont {Wang}, \citenamefont
  {Lehnen},\ and\ \citenamefont {Prosandeev}}]{kle00}%
  \BibitemOpen
  \bibfield  {author} {\bibinfo {author} {\bibfnamefont {W.}~\bibnamefont
  {Kleemann}}, \bibinfo {author} {\bibfnamefont {J.}~\bibnamefont {Dec}},
  \bibinfo {author} {\bibfnamefont {Y.~G.}\ \bibnamefont {Wang}}, \bibinfo
  {author} {\bibfnamefont {P.}~\bibnamefont {Lehnen}}, \ and\ \bibinfo {author}
  {\bibfnamefont {S.~A.}\ \bibnamefont {Prosandeev}},\ }\href@noop {}
  {\bibfield  {journal} {\bibinfo  {journal} {J. Phys. Chem. Solids},\ }\textbf
  {\bibinfo {volume} {61}},\ \bibinfo {pages} {167} (\bibinfo {year}
  {2000})}\BibitemShut {NoStop}%
\bibitem [{\citenamefont {Bednorz}\ and\ \citenamefont
  {M{\"u}ller}(1984)}]{bed84}%
  \BibitemOpen
  \bibfield  {author} {\bibinfo {author} {\bibfnamefont {J.~G.}\ \bibnamefont
  {Bednorz}}\ and\ \bibinfo {author} {\bibfnamefont {K.~A.}\ \bibnamefont
  {M{\"u}ller}},\ }\href@noop {} {\bibfield  {journal} {\bibinfo  {journal}
  {Phys. Rev. Lett.},\ }\textbf {\bibinfo {volume} {52}},\ \bibinfo {pages}
  {2289} (\bibinfo {year} {1984})}\BibitemShut {NoStop}%
\bibitem [{\citenamefont {Geneste}\ and\ \citenamefont {Kiat}(2008)}]{gen08}%
  \BibitemOpen
  \bibfield  {author} {\bibinfo {author} {\bibfnamefont {G.}~\bibnamefont
  {Geneste}}\ and\ \bibinfo {author} {\bibfnamefont {J.-M.}\ \bibnamefont
  {Kiat}},\ }\href@noop {} {\bibfield  {journal} {\bibinfo  {journal} {Phys.
  Rev. B},\ }\textbf {\bibinfo {volume} {77}},\ \bibinfo {pages} {174101}
  (\bibinfo {year} {2008})}\BibitemShut {NoStop}%
\bibitem [{\citenamefont {Carpenter}\ \emph {et~al.}(2006)\citenamefont
  {Carpenter}, \citenamefont {Howard}, \citenamefont {Knight},\ and\
  \citenamefont {Zhang}}]{car06}%
  \BibitemOpen
  \bibfield  {author} {\bibinfo {author} {\bibfnamefont {M.~A.}\ \bibnamefont
  {Carpenter}}, \bibinfo {author} {\bibfnamefont {C.~J.}\ \bibnamefont
  {Howard}}, \bibinfo {author} {\bibfnamefont {K.~S.}\ \bibnamefont {Knight}},
  \ and\ \bibinfo {author} {\bibfnamefont {Z.}~\bibnamefont {Zhang}},\
  }\href@noop {} {\bibfield  {journal} {\bibinfo  {journal} {J. Phys-Condensed
  Matter},\ }\textbf {\bibinfo {volume} {18}},\ \bibinfo {pages} {10725}
  (\bibinfo {year} {2006})}\BibitemShut {NoStop}%
\bibitem [{\citenamefont {Woodward}\ \emph {et~al.}(2006)\citenamefont
  {Woodward}, \citenamefont {Wise}, \citenamefont {Lee},\ and\ \citenamefont
  {Reaney}}]{woo06}%
  \BibitemOpen
  \bibfield  {author} {\bibinfo {author} {\bibfnamefont {D.~I.}\ \bibnamefont
  {Woodward}}, \bibinfo {author} {\bibfnamefont {P.~L.}\ \bibnamefont {Wise}},
  \bibinfo {author} {\bibfnamefont {W.~E.}\ \bibnamefont {Lee}}, \ and\
  \bibinfo {author} {\bibfnamefont {I.~M.}\ \bibnamefont {Reaney}},\
  }\href@noop {} {\bibfield  {journal} {\bibinfo  {journal} {J. Phys-Condensed
  Matter},\ }\textbf {\bibinfo {volume} {18}},\ \bibinfo {pages} {2401}
  (\bibinfo {year} {2006})}\BibitemShut {NoStop}%
\bibitem [{\citenamefont {Hui}\ \emph {et~al.}(2007)\citenamefont {Hui},
  \citenamefont {Dove}, \citenamefont {Tucker}, \citenamefont {Redfern},\ and\
  \citenamefont {Keen}}]{hui07}%
  \BibitemOpen
  \bibfield  {author} {\bibinfo {author} {\bibfnamefont {Q.}~\bibnamefont
  {Hui}}, \bibinfo {author} {\bibfnamefont {M.~T.}\ \bibnamefont {Dove}},
  \bibinfo {author} {\bibfnamefont {M.~G.}\ \bibnamefont {Tucker}}, \bibinfo
  {author} {\bibfnamefont {S.~A.~T.}\ \bibnamefont {Redfern}}, \ and\ \bibinfo
  {author} {\bibfnamefont {D.~A.}\ \bibnamefont {Keen}},\ }\href@noop {}
  {\bibfield  {journal} {\bibinfo  {journal} {J. Phys-Condensed Matter},\
  }\textbf {\bibinfo {volume} {19}},\ \bibinfo {pages} {335214} (\bibinfo
  {year} {2007})}\BibitemShut {NoStop}%
\bibitem [{\citenamefont {Kresse}\ and\ \citenamefont {Hafner}(1993)}]{vasp1}%
  \BibitemOpen
  \bibfield  {author} {\bibinfo {author} {\bibfnamefont {G.}~\bibnamefont
  {Kresse}}\ and\ \bibinfo {author} {\bibfnamefont {J.}~\bibnamefont
  {Hafner}},\ }\href@noop {} {\bibfield  {journal} {\bibinfo  {journal} {Phys.
  Rev. B},\ }\textbf {\bibinfo {volume} {47}},\ \bibinfo {pages} {R558}
  (\bibinfo {year} {1993})}\BibitemShut {NoStop}%
\bibitem [{\citenamefont {Kresse}\ and\ \citenamefont
  {Furthm{\"u}ller}(1996)}]{vasp2}%
  \BibitemOpen
  \bibfield  {author} {\bibinfo {author} {\bibfnamefont {G.}~\bibnamefont
  {Kresse}}\ and\ \bibinfo {author} {\bibfnamefont {J.}~\bibnamefont
  {Furthm{\"u}ller}},\ }\href@noop {} {\bibfield  {journal} {\bibinfo
  {journal} {Phys. Rev. B},\ }\textbf {\bibinfo {volume} {54}},\ \bibinfo
  {pages} {11169} (\bibinfo {year} {1996})}\BibitemShut {NoStop}%
\bibitem [{\citenamefont {Bl{\"o}chl}(1994)}]{paw}%
  \BibitemOpen
  \bibfield  {author} {\bibinfo {author} {\bibfnamefont {P.~E.}\ \bibnamefont
  {Bl{\"o}chl}},\ }\href@noop {} {\bibfield  {journal} {\bibinfo  {journal}
  {Phys. Rev. B},\ }\textbf {\bibinfo {volume} {50}},\ \bibinfo {pages} {17953}
  (\bibinfo {year} {1994})}\BibitemShut {NoStop}%
\bibitem [{\citenamefont {Perdew}\ \emph {et~al.}(1996)\citenamefont {Perdew},
  \citenamefont {Burke},\ and\ \citenamefont {Ernzerhof}}]{pbe96}%
  \BibitemOpen
  \bibfield  {author} {\bibinfo {author} {\bibfnamefont {J.~P.}\ \bibnamefont
  {Perdew}}, \bibinfo {author} {\bibfnamefont {K.}~\bibnamefont {Burke}}, \
  and\ \bibinfo {author} {\bibfnamefont {M.}~\bibnamefont {Ernzerhof}},\
  }\href@noop {} {\bibfield  {journal} {\bibinfo  {journal} {Phys. Rev.
  Lett.},\ }\textbf {\bibinfo {volume} {77}},\ \bibinfo {pages} {3865}
  (\bibinfo {year} {1996})}\BibitemShut {NoStop}%
\bibitem [{\citenamefont {Gonze}\ \emph {et~al.}(2002)\citenamefont {Gonze},
  \citenamefont {Beuken}, \citenamefont {Caracas}, \citenamefont {Detraux},
  \citenamefont {Fuchs}, \citenamefont {Rignanese}, \citenamefont {Sindic},
  \citenamefont {Verstraete}, \citenamefont {Zerah}, \citenamefont {Jollet},
  \citenamefont {Torrent}, \citenamefont {Roy}, \citenamefont {Mikami},
  \citenamefont {Ghosez}, \citenamefont {Raty},\ and\ \citenamefont
  {Allan}}]{abinit}%
  \BibitemOpen
  \bibfield  {author} {\bibinfo {author} {\bibfnamefont {X.}~\bibnamefont
  {Gonze}}, \bibinfo {author} {\bibfnamefont {J.-M.}\ \bibnamefont {Beuken}},
  \bibinfo {author} {\bibfnamefont {R.}~\bibnamefont {Caracas}}, \bibinfo
  {author} {\bibfnamefont {F.}~\bibnamefont {Detraux}}, \bibinfo {author}
  {\bibfnamefont {M.}~\bibnamefont {Fuchs}}, \bibinfo {author} {\bibfnamefont
  {G.-M.}\ \bibnamefont {Rignanese}}, \bibinfo {author} {\bibfnamefont
  {L.}~\bibnamefont {Sindic}}, \bibinfo {author} {\bibfnamefont
  {M.}~\bibnamefont {Verstraete}}, \bibinfo {author} {\bibfnamefont
  {G.}~\bibnamefont {Zerah}}, \bibinfo {author} {\bibfnamefont
  {F.}~\bibnamefont {Jollet}}, \bibinfo {author} {\bibfnamefont
  {M.}~\bibnamefont {Torrent}}, \bibinfo {author} {\bibfnamefont
  {A.}~\bibnamefont {Roy}}, \bibinfo {author} {\bibfnamefont {M.}~\bibnamefont
  {Mikami}}, \bibinfo {author} {\bibfnamefont {P.}~\bibnamefont {Ghosez}},
  \bibinfo {author} {\bibfnamefont {J.-Y.}\ \bibnamefont {Raty}}, \ and\
  \bibinfo {author} {\bibfnamefont {D.}~\bibnamefont {Allan}},\ }\href@noop {}
  {\bibfield  {journal} {\bibinfo  {journal} {Comp. Mater. Sci.},\ }\textbf
  {\bibinfo {volume} {25}},\ \bibinfo {pages} {478} (\bibinfo {year}
  {2002})}\BibitemShut {NoStop}%
\bibitem [{\citenamefont {Fuchs}\ and\ \citenamefont
  {Scheffler}(1999)}]{fhi99}%
  \BibitemOpen
  \bibfield  {author} {\bibinfo {author} {\bibfnamefont {M.}~\bibnamefont
  {Fuchs}}\ and\ \bibinfo {author} {\bibfnamefont {M.}~\bibnamefont
  {Scheffler}},\ }\href@noop {} {\bibfield  {journal} {\bibinfo  {journal}
  {Comput. Phys. Commun.},\ }\textbf {\bibinfo {volume} {119}},\ \bibinfo
  {pages} {67} (\bibinfo {year} {1999})}\BibitemShut {NoStop}%
\bibitem [{\citenamefont {M{\"u}ller}\ and\ \citenamefont
  {Burkard}(1979)}]{mul79}%
  \BibitemOpen
  \bibfield  {author} {\bibinfo {author} {\bibfnamefont {K.~A.}\ \bibnamefont
  {M{\"u}ller}}\ and\ \bibinfo {author} {\bibfnamefont {H.}~\bibnamefont
  {Burkard}},\ }\href@noop {} {\bibfield  {journal} {\bibinfo  {journal} {Phys.
  Rev. B},\ }\textbf {\bibinfo {volume} {19}},\ \bibinfo {pages} {3593}
  (\bibinfo {year} {1979})}\BibitemShut {NoStop}%
\bibitem [{\citenamefont {Zhong}\ and\ \citenamefont
  {Vanderbilt}(1996)}]{zho96}%
  \BibitemOpen
  \bibfield  {author} {\bibinfo {author} {\bibfnamefont {W.}~\bibnamefont
  {Zhong}}\ and\ \bibinfo {author} {\bibfnamefont {D.}~\bibnamefont
  {Vanderbilt}},\ }\href@noop {} {\bibfield  {journal} {\bibinfo  {journal}
  {Phys. Rev. B},\ }\textbf {\bibinfo {volume} {53}},\ \bibinfo {pages} {5047}
  (\bibinfo {year} {1996})}\BibitemShut {NoStop}%
\bibitem [{\citenamefont {Glazer}(1972)}]{gla72}%
  \BibitemOpen
  \bibfield  {author} {\bibinfo {author} {\bibfnamefont {A.}~\bibnamefont
  {Glazer}},\ }\href@noop {} {\bibfield  {journal} {\bibinfo  {journal} {Acta
  Crystallogr. B},\ }\textbf {\bibinfo {volume} {28}},\ \bibinfo {pages} {3384}
  (\bibinfo {year} {1972})}\BibitemShut {NoStop}%
\bibitem [{\citenamefont {Halilov}\ \emph {et~al.}(2002)\citenamefont
  {Halilov}, \citenamefont {Fornari},\ and\ \citenamefont {Singh}}]{hal02}%
  \BibitemOpen
  \bibfield  {author} {\bibinfo {author} {\bibfnamefont {S.~V.}\ \bibnamefont
  {Halilov}}, \bibinfo {author} {\bibfnamefont {M.}~\bibnamefont {Fornari}}, \
  and\ \bibinfo {author} {\bibfnamefont {D.~J.}\ \bibnamefont {Singh}},\
  }\href@noop {} {\bibfield  {journal} {\bibinfo  {journal} {Appl. Phys.
  Lett.},\ }\textbf {\bibinfo {volume} {81}},\ \bibinfo {pages} {3443}
  (\bibinfo {year} {2002})}\BibitemShut {NoStop}%
\bibitem [{\citenamefont {Sai}\ and\ \citenamefont {Vanderbilt}(2000)}]{sai00}%
  \BibitemOpen
  \bibfield  {author} {\bibinfo {author} {\bibfnamefont {N.}~\bibnamefont
  {Sai}}\ and\ \bibinfo {author} {\bibfnamefont {D.}~\bibnamefont
  {Vanderbilt}},\ }\href@noop {} {\bibfield  {journal} {\bibinfo  {journal}
  {Phys. Rev. B},\ }\textbf {\bibinfo {volume} {62}},\ \bibinfo {pages} {13942}
  (\bibinfo {year} {2000})}\BibitemShut {NoStop}%
\bibitem [{\citenamefont {Shannon}(1976)}]{sha76}%
  \BibitemOpen
  \bibfield  {author} {\bibinfo {author} {\bibfnamefont {R.}~\bibnamefont
  {Shannon}},\ }\href@noop {} {\bibfield  {journal} {\bibinfo  {journal} {Acta
  Crystallogr.},\ }\textbf {\bibinfo {volume} {A32}},\ \bibinfo {pages} {751}
  (\bibinfo {year} {1976})}\BibitemShut {NoStop}%
\bibitem [{\citenamefont {Eklund}\ \emph {et~al.}(2009)\citenamefont {Eklund},
  \citenamefont {Fennie},\ and\ \citenamefont {Rabe}}]{ekl09}%
  \BibitemOpen
  \bibfield  {author} {\bibinfo {author} {\bibfnamefont {C.-J.}\ \bibnamefont
  {Eklund}}, \bibinfo {author} {\bibfnamefont {C.~J.}\ \bibnamefont {Fennie}},
  \ and\ \bibinfo {author} {\bibfnamefont {K.~M.}\ \bibnamefont {Rabe}},\
  }\href@noop {} {\bibfield  {journal} {\bibinfo  {journal} {Phys. Rev. B},\
  }\textbf {\bibinfo {volume} {79}},\ \bibinfo {pages} {220101(R)} (\bibinfo
  {year} {2009})}\BibitemShut {NoStop}%
\bibitem [{\citenamefont {Haeni}\ \emph {et~al.}(2004)\citenamefont {Haeni},
  \citenamefont {Irvin}, \citenamefont {Chang}, \citenamefont {Uecker},
  \citenamefont {Li}, \citenamefont {Choudhury}, \citenamefont {Tian},
  \citenamefont {Hawley}, \citenamefont {Craigo}, \citenamefont {Tagantsev},
  \citenamefont {Pan}, \citenamefont {Streiffer}, \citenamefont {Chen},
  \citenamefont {Kirchoefer}, \citenamefont {Levy},\ and\ \citenamefont
  {Schlom}}]{hae04}%
  \BibitemOpen
  \bibfield  {author} {\bibinfo {author} {\bibfnamefont {J.~H.}\ \bibnamefont
  {Haeni}}, \bibinfo {author} {\bibfnamefont {P.}~\bibnamefont {Irvin}},
  \bibinfo {author} {\bibfnamefont {W.}~\bibnamefont {Chang}}, \bibinfo
  {author} {\bibfnamefont {R.}~\bibnamefont {Uecker}}, \bibinfo {author}
  {\bibfnamefont {P.~R. Y.~L.}\ \bibnamefont {Li}}, \bibinfo {author}
  {\bibfnamefont {S.}~\bibnamefont {Choudhury}}, \bibinfo {author}
  {\bibfnamefont {W.}~\bibnamefont {Tian}}, \bibinfo {author} {\bibfnamefont
  {M.~E.}\ \bibnamefont {Hawley}}, \bibinfo {author} {\bibfnamefont
  {B.}~\bibnamefont {Craigo}}, \bibinfo {author} {\bibfnamefont {A.~K.}\
  \bibnamefont {Tagantsev}}, \bibinfo {author} {\bibfnamefont {X.~Q.}\
  \bibnamefont {Pan}}, \bibinfo {author} {\bibfnamefont {S.~K.}\ \bibnamefont
  {Streiffer}}, \bibinfo {author} {\bibfnamefont {L.~Q.}\ \bibnamefont {Chen}},
  \bibinfo {author} {\bibfnamefont {S.~W.}\ \bibnamefont {Kirchoefer}},
  \bibinfo {author} {\bibfnamefont {J.}~\bibnamefont {Levy}}, \ and\ \bibinfo
  {author} {\bibfnamefont {D.~G.}\ \bibnamefont {Schlom}},\ }\href@noop {}
  {\bibfield  {journal} {\bibinfo  {journal} {Nature},\ }\textbf {\bibinfo
  {volume} {430}},\ \bibinfo {pages} {758} (\bibinfo {year}
  {2004})}\BibitemShut {NoStop}%
\end{thebibliography}%
%====================================================================
\end{document}